\documentclass[8.5pt,twoside,twocolumn]{article}
\usepackage[super,sort&compress,comma]{natbib} 
\usepackage{mhchem}
\usepackage{times,mathptm}
\usepackage{sectsty}
\usepackage{balance} 

\usepackage{graphicx} 
\usepackage{lastpage}
\usepackage[format=plain,justification=raggedright,singlelinecheck=false,font=small,labelfont=bf,labelsep=space]{caption} 
\usepackage{fancyhdr}
\usepackage{amssymb,amsmath}
\usepackage{booktabs}

\begin{document}

\thispagestyle{plain}
\renewcommand{\thefootnote}{\fnsymbol{footnote}}
\renewcommand\footnoterule{\vspace*{1pt}%
\hrule width 3.4in height 0.4pt \vspace*{5pt}}

\makeatletter 
\renewcommand\@biblabel[1]{#1}            
\renewcommand\@makefntext[1]%
{\noindent\makebox[0pt][r]{\@thefnmark\,}#1}
\makeatother 
\renewcommand{\figurename}{\small{Fig.}~}
\sectionfont{\large}
\subsectionfont{\normalsize} 

\fancyfoot{}
\fancyfoot[RO]{\footnotesize{\sffamily{1--\pageref{LastPage} ~\textbar  \hspace{2pt}\thepage}}}
\fancyfoot[LE]{\footnotesize{\sffamily{\thepage~\textbar\hspace{4.4cm} 1--\pageref{LastPage}}}}
\fancyhead{}
\renewcommand{\headrulewidth}{1pt} 
\renewcommand{\footrulewidth}{1pt}
\setlength{\arrayrulewidth}{1pt}
\setlength{\columnsep}{6.5mm}
\setlength\bibsep{1pt}

\twocolumn[
  \begin{@twocolumnfalse}
\noindent\LARGE{\textbf{Measuring individual overpotentials in an operating solid-oxide electrochemical cell}}
\vspace{0.6cm}

\noindent\large{\textbf{Farid El~Gabaly,$^{\ast}$\textit{$^{a}$} Michael Grass,\textit{$^{b}$} Anthony H. McDaniel,\textit{$^{a}$} Roger L. Farrow,\textit{$^{a}$} Mark A. Linne,$^{\dag}$\textit{$^{a}$} Zahid Hussain,\textit{$^{b}$} Hendrik Bluhm,\textit{$^{c}$} Zhi Liu,$^{\ast}$\textit{$^{b}$} and
Kevin F. McCarty\textit{$^{a}$}}}\vspace{0.5cm}

\noindent\textit{\small{\textbf{Received Xth XXXXXXXXXX 20XX, Accepted Xth XXXXXXXXX 20XX\newline
First published on the web Xth XXXXXXXXXX 200X}}}

\noindent \textbf{\small{DOI: 10.1039/b000000x}}
\vspace{0.6cm}

\noindent \normalsize{We use photo-electrons as a non-contact probe to measure local electrical potentials in a solid-oxide electrochemical cell. We characterize the cell \textit{in~operando} at near-ambient pressure using spatially-resolved X-ray photoemission spectroscopy. The overpotentials at the interfaces between the Ni and Pt electrodes and the yttria-stabilized zirconia (YSZ) electrolyte are directly measured. The method is validated using electrochemical impedance spectroscopy. Using the overpotentials, which characterize the cell's inefficiencies, we compare without ambiguity the electro-catalytic efficiencies of Ni and Pt, finding that on Ni H$_2$O splitting proceeds more rapidly than H$_2$ oxidation ,while on Pt, H$_2$ oxidation proceeds more rapidly than H$_2$O splitting.}
\vspace{0.5cm}
 \end{@twocolumnfalse}
  ]

\footnotetext{\textit{$^{a}$~Sandia National Laboratories, CA 94550, USA. Fax: 01 925 294 3231; Tel: 01 925 294 3441; E-mail: felgaba@sandia.gov}}
\footnotetext{\textit{$^{b}$~Advanced Light Source, Lawrence Berkeley National Laboratory, Berkeley, CA 94720, USA. Fax: 01 510 486 4773; Tel: 01 510 486 2109; E-mail: zliu2@lbl.gov}}
\footnotetext{\textit{$^{c}$~Chemical Sciences Division, Lawrence Berkeley National Laboratory, Berkeley, CA 94720, USA}}

\footnotetext{\dag~Present address: Chalmers University of Technology, Gothenburg, Sweden.}

\section{Introduction}
\label{introduction}
Electrochemical technologies offer very efficient (40\%-95\%) routes to convert and store energy while not introducing carbon-containing species into the atmosphere. Thus, it is widely anticipated that electrochemical technologies will be increasingly used to provide energy that does not contribute to climate change, i.e., carbon-neutral storage and conversion. Batteries, ultra-capacitors, fuel cells and electrolyzers are the most important\cite{winter_what_2004} electrochemical devices used to inter-convert electrical and chemical energy. Fuel cells\cite{haile_fuel_2003,kirubakaran_reviewfuel_2009} and electrolyzers\cite{ni_technological_2008} are closely related: the former converts fuel to electricity and the latter reverses the process. In fact, a single device can accomplish both tasks, depending if it is fed fuel or driven by electricity.

A natural question is why fuel cells and electrolyzers are not widely used even though they were discovered over 150 years ago\cite{grove_xxiv.voltaic_1839,appleby_sir_1990}? Part of the answer may be that fuel cells and electrolyzers have suffered from a disproportionately high fraction of \textit{top-down} research compared to other technologies\cite{andjar_fuel_2009}. That is, the early success at making working devices has led to an emphasis on engineering-based solutions. Innovation resulting from scientific understanding has lagged because essential fundamental knowledge is missing and difficult to obtain. Specifically, which of the basic processes of the charge-transfer reactions\cite{vogler_modelling_2009,goodwin_modeling_2009} limit rates and efficiencies in fuel cells and electrolyzers is still incompletely understood.

\begin{figure}[t!]
 \centerline{\includegraphics[width=0.75\columnwidth]{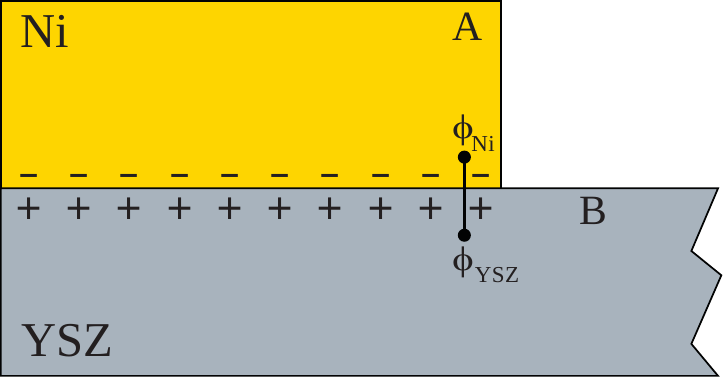}}
  \caption{Schematic of an electrical double-layer at the interface between a metal (Ni) and a solid state electrolyte (YSZ). The inner potentials $\phi$ on either side of the interface are labeled. PES data from representative points A and B are used to determine the overpotential $\eta$ of the interface.}
  \label{fig_double}
\end{figure}

The fundamental electrochemical phenomenon in both fuel cells and electrolyzers is the formation of electrical double-layers across material interfaces, as illustrated in Fig.~\ref{fig_double}. Driven by gradients in their chemical and electrical potentials, charged species (electrons and ions) cross through the double-layers from one phase to another. As reviewed in Section~\ref{potentials}, every double-layer has an associated potential (difference) that is modified by an ``overpotential'' when the current is flowing. In fact, the current through a double-layer depends exponentially on its overpotential\cite{bockris_modern_2000}. An overpotential always decreases the device's power output by reducing the useful potential obtained from a fuel cell and increasing the potential needed to drive an electrolyzer. If the overpotential grows relatively fast with increasing current, the device is inefficient and must be run at low current\cite{foot3} to minimize the losses.

Many theoretical works have addressed the origin and contributions to overpotentials\cite{nrskov_origin_2004,mann_application_2006,bessler_influence_2007}, but experimental approaches to directly measuring overpotentials are scarce because of large challenges. First, overpotentials only exist when the cell is \textit{in operando}. Second, they cannot be measured by a contact probe, the traditional approach to measuring potentials on metals and semiconductors ---~contacting a potential probe to a solid-state electrolyte, which must be electrically insulating, introduces another double-layer, confounding the measurement\cite{gileadi_electrode_1993,bockris_modern_2000}.

\begin{figure}[b!]
  \centerline{\includegraphics[width=0.65\columnwidth]{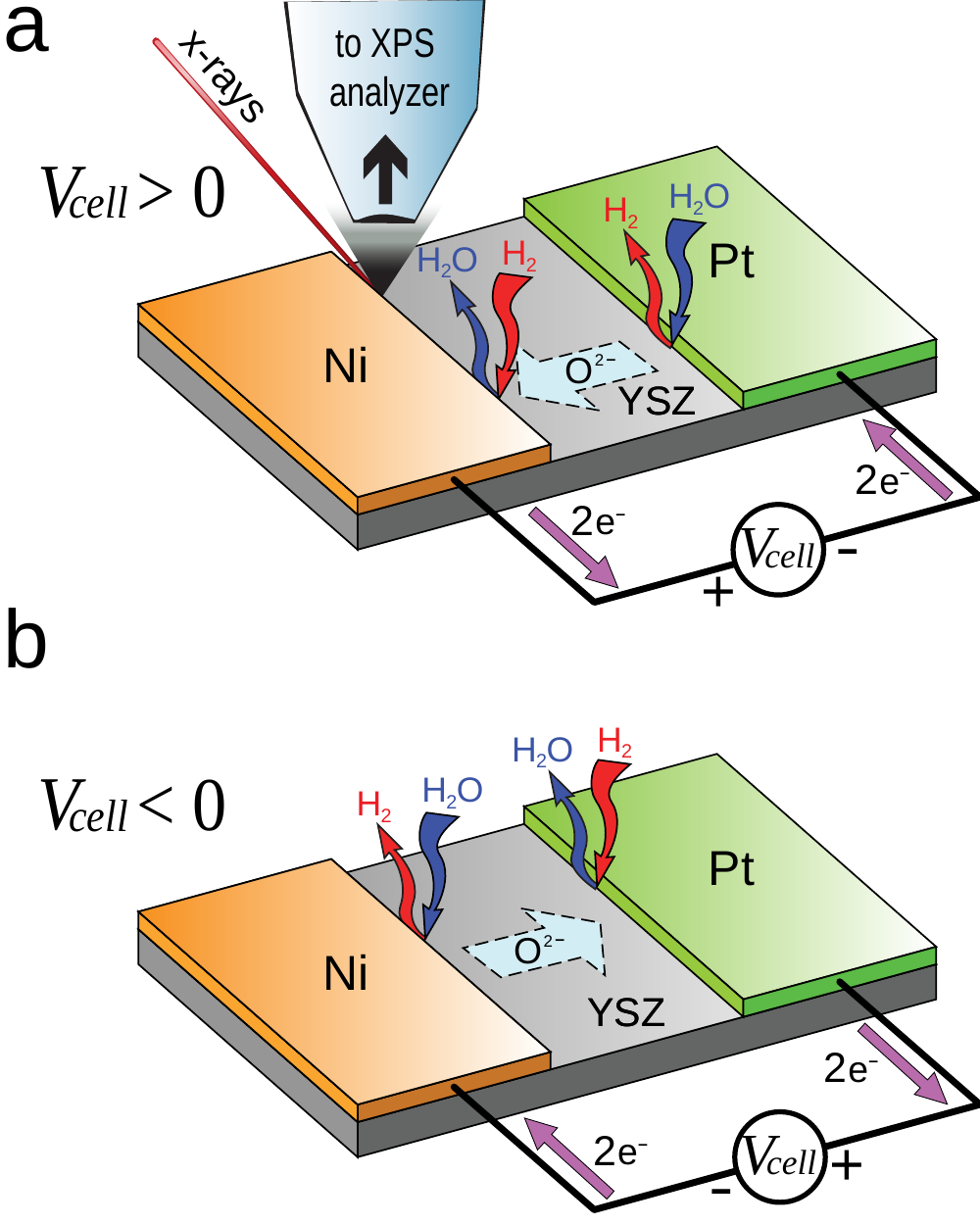}}
  \caption{Schematic of the driven solid-oxide electrochemical cell used in this work. (a) and (b) show the two possible polarities and the gas-phase reactions at each three-phase boundary. (a) also shows the x-ray beam and the collection cone of the spectrometer.}
  \label{fig_sofc}
\end{figure}

Here we use a new method to measure directly all the overpotentials in a working electrochemical device. Using a newly constructed spatially-resolved ambient-pressure photoemission spectroscopy (APXPS) platform\cite{ogletree_differentially_2002,salmeron_ambient_2008,Grass_2010}, we employ photoelectrons as a non-contact probe of local electrical potential\cite{fahlman_electron_1966,siegbahn_method_1982}. We use a driven solid-oxide electrochemical cell (SOEC) as a model system to study both fuel cell (H$_2$ oxidation) and electrolyzer (H$_2$O splitting) reactions on Ni and Pt electrodes (Fig.~\ref{fig_sofc}). Sections~\ref{potentials} and \ref{reactions} describe in detail the potentials and overpotentials expected in this model cell. We then describe the measured overpotentials, which we validate with simultaneous electrochemical impedance spectroscopy (EIS). Analysis of the individual spatially-resolved overpotentials reveals direct information about the charge-transfer reactions.
\section{Experimental}
\label{experimental}
\begin{figure}[t!]
  \centerline{\includegraphics[width=0.65\columnwidth]{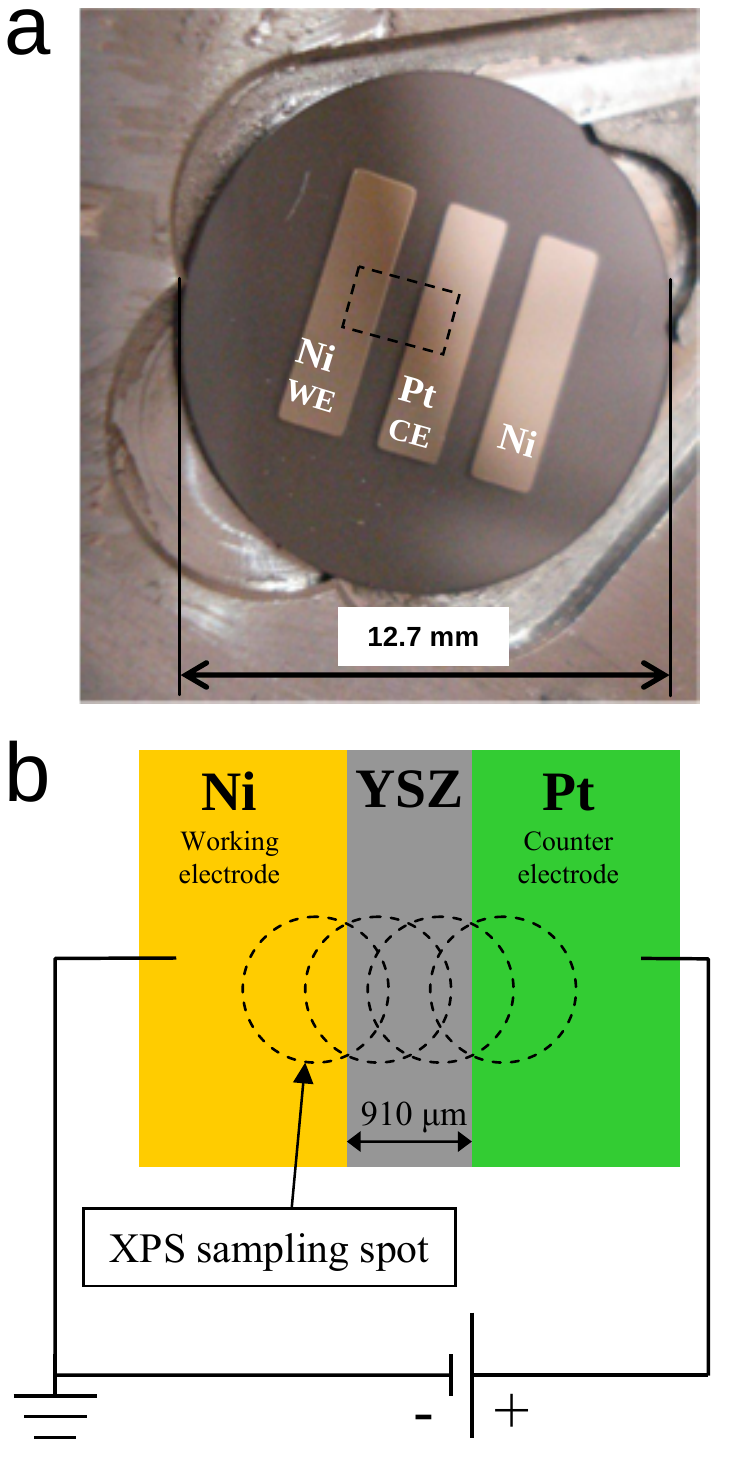}}
  \caption{(a) Photograph of the solid-oxide electrochemical cell. The dashed square region marks the region measured by PES. (b) Schematic (top view) of the measured region, electrode assignment, and cell connections for $V_{cell}<0$. Dashed circles show overlapping regions of PES analysis that span between the two electrodes.}
  \label{fig_photo}
\end{figure}

The solid electrolyte of the SOEC studied was a wafer of single-crystal yttria-stabilized zirconia (YSZ). The Pt (99.99\% pure) and Ni (99.995\% pure) electrodes were fabricated on the YSZ by evaporation from crucibles heated by an electron beam under high vacuum conditions ($10^{-6}$~Torr). The electrode shape was controlled using a stainless-steel shadow mask positioned in front of the YSZ substrate. Figures \ref{fig_sofc} and \ref{fig_photo}a show a schematic and a photograph, respectively, of the SOEC.

A specially developed holder was used to electrically contact the SOEC while at high temperature in the H$_2$O/H$_2$ ambient. The holder uses spring-loaded probes to provide reliable electrical contact to the thin-film electrodes and is equipped with a ceramic button-heater compatible with oxidizing atmospheres\cite{Whaley_2010}. A computer-based Gamry potentiostat (model PCI4-750) was used to bias the SOEC and perform standard electrochemical tests.

\begin{figure*}[t!]
  \centerline{\includegraphics[width=0.8\textwidth]{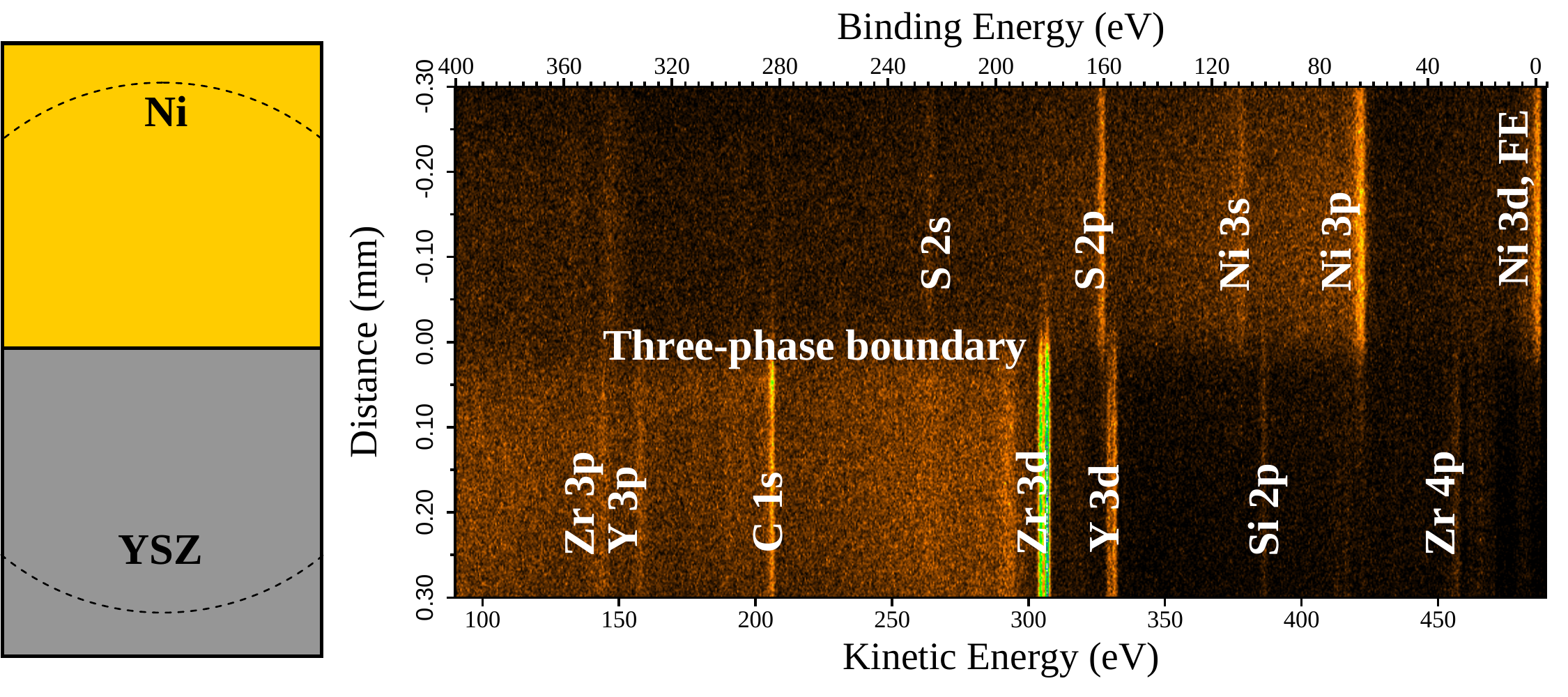}}
  \caption{Left panel: top-view schematic of the SOEC Ni/YSZ interface measured in the ambient-pressure PES image (right). The dashed lines shows the field-of-view ($\sim$0.6~mm). Right panel: the photoelectron binding energy versus real-space distance around the cell's three-phase boundary during operation at zero bias. Core-level PES peaks of Ni, YSZ, their impurities and the Ni Fermi edge (FE) are labeled. Intensities (counts) are displayed using a false-color scale. The binding energy scale is referenced to the Fermi edge of the grounded Ni electrode.}
  \label{fig_survey}
\end{figure*}
Once the cell was positioned in front of the collection cone of the APXPS spectrometer, 150~mTorr H$_2$O was introduced into the chamber, followed by 150~mTorr H$_2$. After stabilizing the pressure, the sample was heated from room temperature to about 700~C over the course of 1 hour\cite{foot5}. These pressures and temperatures were sufficient to produce cell currents up to 200~{$\mu$}A. Temperatures were measured by a two-color pyrometer enabled by blocking the radiation from the heater with a Pt film on the backside of the YSZ wafer. Secondary temperature measurements were obtained by measuring the bulk conductance of YSZ between the electrodes using EIS at zero applied bias. The Ni electrode was assigned to be the working electrode (WE) and was electrically connected to the spectrometer ground. This ``Fermi-edge coupling'' aligns the Fermi edge of the Ni and the spectrometer. The Pt electrode, assigned as the counter electrode (CE), is biased relative to the grounded Ni electrode. The electrochemical measurements were taken in a two probe configuration (Fig.~\ref{fig_photo}b).

APXPS\cite{ogletree_differentially_2002,salmeron_ambient_2008} measurements were performed at the Advanced Light Source (LBNL, Berkeley) beamline 9.3.2. The end-station is equipped with a Scienta R4000 HiPP ambient-pressure electron spectrometer. By means of a series of electron lenses and differentially pumped stages, the electron analyzer side is maintained under high vacuum conditions. The spectrometer was optimized in this project to perform 1D spatially-resolved APXPS\cite{Grass_2010}. The photo-electrons are dispersed by their kinetic energy along the x-axis of the spectrometer's two-dimensional detector. Along the detector's y-axis, the electrons are mapped according to the position that they were emitted from the sample (see Fig.~\ref{fig_survey}). Thus, the spectrometer spectrally resolves the electrons with simultaneous 1D spatial resolution. The field-of-view for the spatial resolution dimension is $\sim$0.6~mm and the resolution is $\sim$1~{$\mu$}m. To analyze larger areas, the sample holder was translated with high precision using a 4-axis manipulator, as illustrated in Fig.~\ref{fig_photo}b. The resulting data was analyzed by dividing each image into 5 or 10~{$\mu$}m thick slices along the spatial resolution axis. By integrating the counts of every slice along this axis, a standard XPS spectrum of counts vs. kinetic energy was obtained. The beamline energy resolution, which determines the XPS peak widths, is 160~meV at the photon energy used (490~eV). 

For obtaining the local electrical potentials of the cell, APXPS spectra were obtained at several values of applied bias. The potentials from the YSZ electrolyte and the Ni and Pt electrodes were measured using the Ni~3p$_{3/2}$, Zr~3d$_{5/2}$ and Pt~4f$_{7/2}$ XPS core levels, respectively. These element-specific core levels have sharp XPS peaks and have relatively high cross-sections for the exciting x-rays~(490~eV). The peaks were background corrected and fitted to determine their kinetic energies. At each discrete analysis location, the peak positions with and without applied bias were subtracted (to obtain ${}^{\rm{bias}}\Delta^{\rm{eq}}\phi$, as explained in detail in Appendix~\ref{potentials_XPS}). The 910~$\mu$m-wide region of YSZ electrolyte (see Fig.\ref{fig_photo}b) between the electrodes was analyzed using 182 Zr~3d$_{5/2}$ spectra separated by 5~{$\mu$}m steps.

\section{Double-layer potentials in a driven SOEC}
\label{potentials}
Since driven (electrolyzers) cells are much less discussed than fuel cells in the literature, in this section we review their operation. We then formally define the different types of potentials present in the SOEC. With no applied bias, the SOEC materials are in equilibrium with the gas phase. Furthermore, the gas phase is in equilibrium with itself, so there is no chemical energy that can be extracted. Applying a potential (bias) between the electrodes perturbs this equilibrium. For the Ni/YSZ/Pt cell shown in Fig.~\ref{fig_sofc}, the metallic electrodes then charge by accumulating electrons or holes, depending on the polarity, at their surfaces. These charged electrodes induce a layer of net (ionic) charge of opposite sign in the electrolyte region adjacent to the electrode, forming an electrical double-layer there. The mobile charges in YSZ are O$^{2-}$ ions\cite{chiang_physical_1997}, so the YSZ under the positive electrode has an O$^{2-}$ excess and the YSZ under the negative electrode has an O$^{2-}$ deficiency, or equivalently, an excess of oxygen vacancies (V$^{\bullet\bullet}_{\rm{O}}$). The electrolyte charge layer is expected to be only a few nanometers thick\cite{bard_electrochemical_2001}. Thus, the double-layer is the response of the electrolyte to the electric field created between the electrodes. The charge at the electrolyte interfaces effectively screens the penetrating electric field and maintains the bulk electrolyte at an almost constant potential, reducing its overall electrostatic energy.

The potential of the charged layer is called the outer (Volta) potential $\psi$ and is of coulombic nature\cite{trasatti_interphases_1986,trasatti_absolute_1986}. The total (inner or Galvani) potential $\phi$ is the sum of the outer potential $\psi$ and the surface potential $\chi$, which is of dipolar nature\cite{trasatti_interphases_1986,trasatti_absolute_1986}. The difference between the inner potentials of the electrode/electrolyte adjacent to their interface is defined as the total double-layer potential and is written for the Ni/YSZ interface as: 
\begin{equation}
{}^{^{\rm{Ni}}}\Delta^{^{\rm{YSZ}}}\phi = {}^{^{\rm{Ni}}}\Delta^{^{\rm{YSZ}}}\chi + {}^{^{\rm{Ni}}}\Delta^{^{\rm{YSZ}}}\psi,
\label{eq:total_pot}
\end{equation}
where $\Delta$ is the difference operator and the $\rm{Ni}$/$\rm{YSZ}$ superscripts represent the Ni-electrode and the YSZ-electrolyte sides of the electrical double-layer, respectively.

The total interface overpotential is defined\cite{bockris_modern_2000} as the change of the inner potential difference between equilibrium (zero external bias) and non-equilibrium (non-zero bias):
\begin{equation}
\eta_{_{\rm{Ni-YSZ}}} = {}^{^{\rm{Ni}}}\Delta^{^{\rm{YSZ}}}\phi_{{\rm{bias}}} - {}^{^{\rm{Ni}}}\Delta^{^{\rm{YSZ}}}\phi_{{\rm{eq}}},
\end{equation}
or equivalently, 
\begin{equation} 
\eta_{_{\rm{Ni-YSZ}}} = {}^{\rm{bias}}\Delta^{\rm{eq}}\phi^{Ni} - {}^{\rm{bias}}\Delta^{\rm{eq}}\phi^{\rm{YSZ}},
\label{eq:eta_2}
\end{equation}  
for the Ni/YSZ interface, where $\phi_{\rm{bias}}$ and $\phi_{\rm{eq}}$ are the biased and the equilibrium inner potentials of the interface, respectively, and $\phi^{\rm{Ni}}$ and $\phi^{\rm{YSZ}}$ are the inner potentials at the Ni and YSZ sides of the double-layer, respectively. These total double-layer overpotentials of a driven SOEC have several contributions\cite{gileadi_electrode_1993}: 
\begin{equation}
\begin{array}{l}
\eta_{_{\rm{Ni-YSZ}}} =\eta_{\rm{bias,_{Ni-YSZ}}} - \eta_{\rm{ac,_{Ni-YSZ}}} - \eta_{\rm{c,_{Ni-YSZ}}} \\
\eta_{_{\rm{YSZ-Pt}}} =\eta_{\rm{bias,_{YSZ-Pt}}} - \eta_{\rm{ac,_{YSZ-Pt}}} - \eta_{\rm{c,_{YSZ-Pt}}}
\end{array}
\label{eq:eta_ysz-pt}
\end{equation}
where $\eta_{\rm{bias,_{Ni-YSZ}}}$ and $\eta_{\rm{bias,_{YSZ-Pt}}}$ are the portions of the externally applied bias that creates the double-layers at the electrode/YSZ interfaces, $\eta_{\rm{ac,_{Ni-YSZ}}}$ and $\eta_{\rm{ac,_{YSZ-Pt}}}$ are the reaction activation overpotentials and $\eta_{\rm{c,_{Ni-YSZ}}}$ and $\eta_{\rm{c,_{YSZ-Pt}}}$ are the mass-transfer limitation overpotentials, also called the concentration overpotentials. We note that all the contributions to the overpotential are interrelated\cite{gileadi_electrode_1993} and it is not possible to obtain the total overpotentials from independently calculated contributions. As we next review, only the $\eta_{\rm{bias}}$ contributions favor the electrochemical reactions.

\section{Electrochemical reactions in a driven SOEC}
\label{reactions}
When the driven SOEC is exposed to gas-phase H$_2$ and H$_2$O, charge-transfer reactions start (Fig.~\ref{fig_sofc}). The electrical and chemical driving forces for the charge-transfer reactions arise from the double-layers at the electrode/electrolyte interfaces, which scale as $\eta_{\rm{bias,_{Ni-YSZ}}}$ and $\eta_{\rm{bias,_{YSZ-Pt}}}$. The O$^{2-}$ and V$^{\bullet\bullet}_{\rm{O}}$ excess under the electrodes at the opposite sides of the YSZ electrolyte will drive the adjacent bare YSZ surface out of \textit{chemical} equilibrium with the gas-phase species. Gas/surface reactions then act to return the surface to equilibrium. In addition, the strong electric field produced by the double-layers creates a gradient in the electrochemical free energy of electrons and ionic species involved in the charge-transfer reactions. This gradient causes electrons to cross the electrode/gas interface and the ions to cross the electrolyte/gas interface (in the most general case\cite{adler_factors_2004}).

Since charge-transfer reactions involve gas-phase species, electrons, and O$^{2-}$ ions, they will occur at or close to the three-phase boundary (TPB), where the gas phase, electrolyte and electrode meet, and only where a double-layer exists. The overall charge-transfer reaction,
\begin{equation}
\text{H}_2^{\text{gas}} + \text{O}^{2-}_{\text{YSZ}} \leftrightarrows \text{2e}^-_{\text{electrode}} + \text{H}_2\text{O}^{\text{gas}}, 
 \label{eq:overall_rxn}
\end{equation}
proceeds at the same net rates in the forward and reverse directions at the TPB of the positive and negative electrodes, respectively (see Fig.~\ref{fig_sofc}). 

In response to the charge-transfer reactions, the electrolyte produces a net flux of O$^{2-}$ from the negative to the positive electrode, closing the circuit with the external, fixed-potential power supply. The potential drop arising from the O$^{2-}$ flux across the YSZ electrolyte is a different type of overpotential, since it does not occur at an interface, but though the electrolyte. It has the form $\eta_{\rm{_{R}}}=I_{cell}R_{YSZ}$, where $I_{cell}$ is the total cell current and $R_{YSZ}$ is the electrolyte resistance. 

The Butler-Volmer kinetic relationship,
\begin{equation} 
{i}\propto{i}_{0}{e}^{\eta},
\label{eq:butler_volmer}
\end{equation}
in its simplest form (the high-field approximation\cite{bockris_modern_2000,noren_clarifyingbutler-volmer_2005}), relates the current density $i$ and the total double-layer overpotentials $\eta$. ${i}_{0}$ is the exchange-current density flowing back and forth across the interface when the cell is in equilibrium, and is proportional to the charge-transfer reaction rate. The applied potential difference $V_{cell}$ is distributed across the three overpotentials of the cell:
\begin{equation}
V_{\rm{cell}} = \eta_{_{\rm{Ni-YSZ}}} + \eta_{_{\rm{YSZ-Pt}}} + \eta_{\rm{_{R}}}
\label{eq:V}
\end{equation}
However, the Butler-Volmer relationship tells us that the total overpotentials of the two electrode/electrolyte interfaces, not $V_{cell}$, directly controls the current across the interface. Thus, to balance all the currents through the cell at steady state requires a unique, non-trivial set of overpotentials (Eq.~\ref{eq:V}). By directly measuring these fundamental and highly correlated values, that is, by determining the ``potential landscape,'' SOEC performance can be understood and improved. We next quantify the potential landscape.

\section{Results and discussion}
\label{results}
\begin{figure}[t!]
  \centerline{\includegraphics[width=0.85\columnwidth]{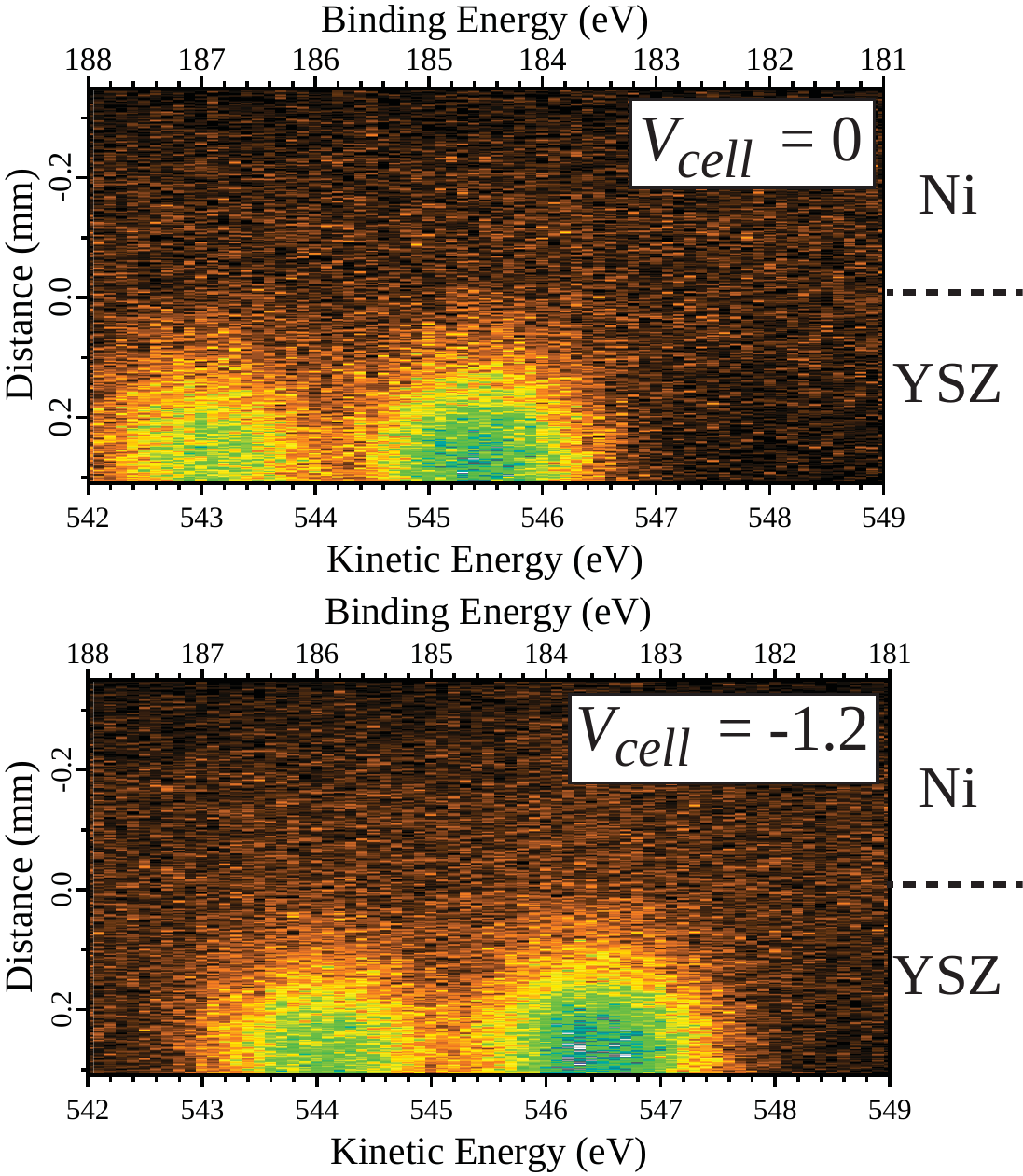}}
  \caption{Example of a rigid shift in the kinetic energy of a XPS peak when bias is applied across the Ni and Pt electrodes. In the image, the Zr~3d doublet ends at the boundary between the YSZ electrolyte and the grounded Ni electrode. The Zr~3d doublet shifts some fraction of the cell bias V$_{cell}$ because the YSZ inner potential $\phi$ changes when a double-layer forms under the Ni electrode. }
  \label{fig_shift}
\end{figure}
In this section we will describe how we have directly measured the three SOEC overpotentials $\eta_{_{\rm{Ni-YSZ}}}$, $\eta_{_{\rm{YSZ-Pt}}}$ and $\eta_{_{\rm{R}}}$ of Eq.~\ref{eq:V} using APXPS. We then discuss the information contained in the overpotentials. Figure \ref{fig_shift} illustrates how the kinetic energy of a Zr XPS peak changes with cell bias. During the SOEC operation, the Pt, Ni, and YSZ surfaces underwent no chemical changes, as revealed through the XPS spectra. That is, the peak shifts with cell bias have no contributions from chemical shifts. Then the kinetic energy changes only arise from changes in the material's inner potential, as shown in Eq.~\ref{eq:delta_inner_apend_V} of the Appendix~\ref{potentials_XPS}. Figure \ref{fig_potentials}a plots the element-specific absolute values of $E_{\rm{K.E.}}^{\rm{bias}} - E_{\rm{K.E.}}^{\rm{eq}} = {}^{\rm{bias}}\Delta^{\rm{eq}}\phi$ from the Ni electrode, across the YSZ electrolyte, to the Pt electrode for $V_{cell}=-1.2$~V. The potentials are constant across the two metal electrodes\cite{foot4}. The standard deviations of the kinetic energy differences with and without -1.2~V bias are $\sigma($Ni~3p$_{3/2})=0.012$~eV, and $\sigma($Pt~4f$_{7/2})=0.002$~eV for Ni and Pt, respectively. These results show the high precision, roughly 10 meV, at which potentials and potential differences can be measured by the APXPS approach. Since the precision depends mainly on the signal-to-noise ratio of the XPS peak, it can be easily improved using greater time averaging, for example. Across the YSZ the potential changes linearly, as documented by the goodness of the least-squares fit to a line $y$=a+b$x$: a$=0.783{\pm}0.003$~eV and b$=5.181{\times}10^{-5}{\pm}2.78{\times}10^{-6}$~{eV/$\mu$m} where $x$ is distance. 

\begin{center}\begin{table*}[!ht] 
{\hfill{}
\begin{tabular}{cccccc}
\toprule
\multicolumn{2}{c}{Potentiostat} &  \multicolumn{3}{c}{XPS} & \multicolumn{1}{c}{$\eta_{_{\rm{R}}}/I_{cell}$}\\
\cmidrule(r){1-2}
\cmidrule(r){3-5}
\cmidrule(r){6-6}
$V_{cell}$(V) & $I_{cell} (\mu$A) & $\eta_{_{\rm{Ni-YSZ}}}$(V)& $\eta_{_{\rm{YSZ-Pt}}}$(V) & $\eta_{_{\rm{R}}}$(V) & $R_{YSZ} (\Omega$)\\
\cmidrule(r){1-2}
\cmidrule(r){3-5}
\cmidrule(r){6-6}
-1.2 & 202 & 0.81 & 0.30 & 0.045 & 223\\
-0.6 & 86  & 0.32 & 0.20 & 0.017 & 198\\
0.6  & 48  & 0.37 & 0.17 & 0.010 & 208\\
1.2  & 32 & 1.02 & 0.15 & 0.006 & 187\\ \bottomrule
\end{tabular}}\hfill{}
\caption{Measured values for the Ni/YSZ/Pt SOEC, obtained from the potentiostat and from the APXPS experiment, as indicated. The rightmost column shows the value of $R_{YSZ}$ obtained by dividing the electrolyte overpotential, $\eta_{_{\rm{R}}}$, by the total cell current, $I_{cell}$.}
\label{table_values}
\end{table*}\end{center}
\begin{figure}[t!]
  \centerline{\includegraphics[width=0.75\columnwidth]{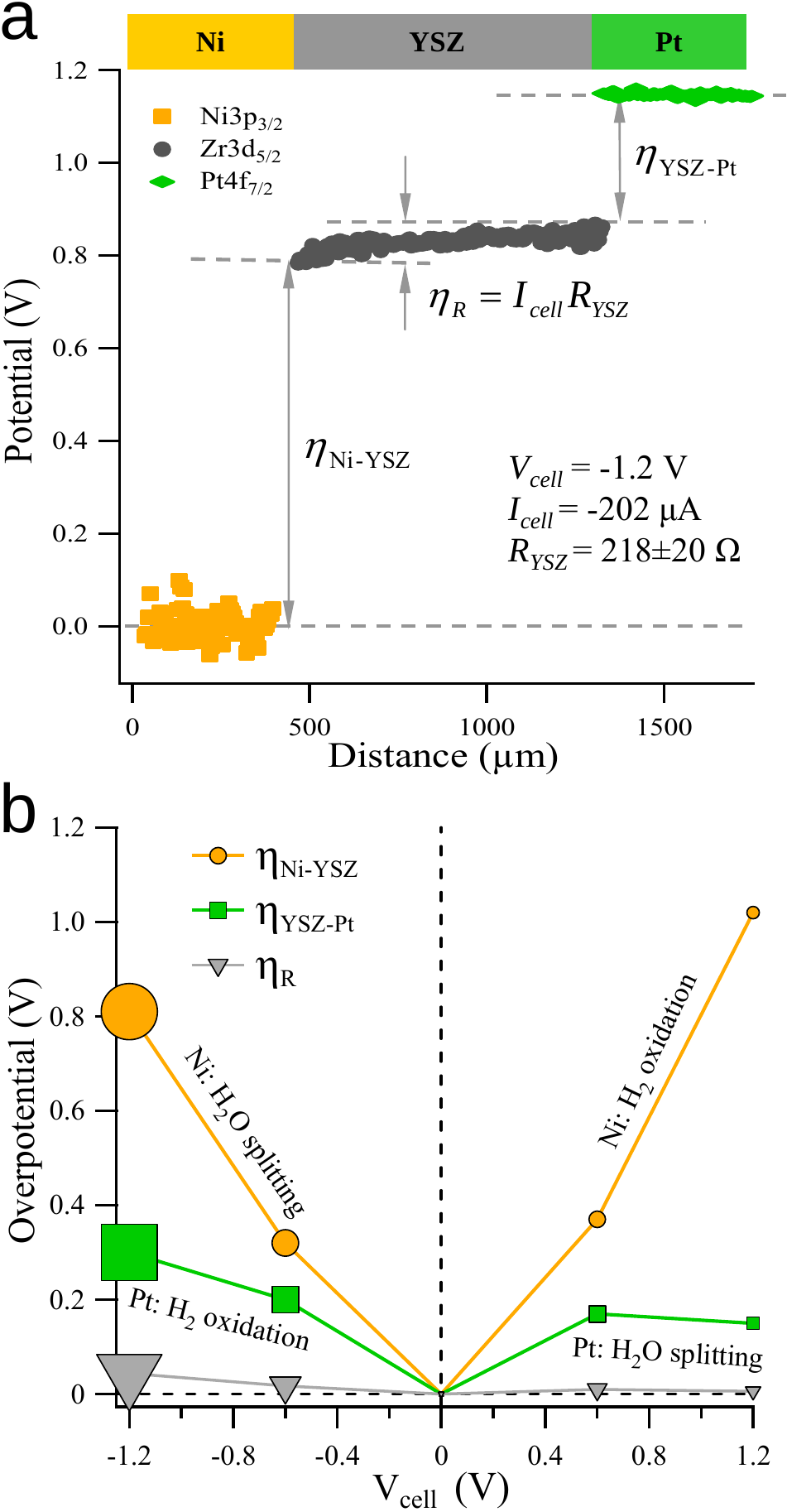}}
  \caption{(a) Potential landscape of the SOEC at $V_{cell} = -1.2 $ V. The labeled overpotentials for the Ni/YSZ and YSZ/Pt interfaces appear as discontinuities in the potential. (b) Measured, absolute values of the three total overpotentials vs. cell bias (see Table~\ref{table_values}). The plot labels give the net reactions taking place at each electrode's TPB for each given condition. The marker size is proportional to the cell current, $I_{cell}$.}
  \label{fig_potentials}
\end{figure}

In Figure \ref{fig_potentials}a the discontinuities in the potentials (i.e., the  ${}^{\rm{bias}}\Delta^{\rm{eq}}\phi$ values) at the electrode/electrolyte boundaries are the overpotentials, as established by Eq.~\ref{eq:eta_2}. Because the electrodes are metals, their electrical potential is constant. Thus, the metal has the same potential at the point probed by XPS (point A in Fig.~\ref{fig_double}) and at its buried interface with the YSZ. We use the 1D spatial resolution to measure the inner potential of the bare YSZ at point B in Fig.~\ref{fig_double}, which is about 10~$\mu$m from the electrode. The YSZ potential at this point and the potential of the YSZ under the electrode differ only by the small ohmic drop through the short YSZ segment. For the operating conditions of our cell, we estimate that this difference introduces an error of less than 1\% of $\eta_{_{\rm{R}}}$ in the double-layer overpotentials.

\begin{figure}[ht!]
  \centerline{\includegraphics[width=0.75\columnwidth]{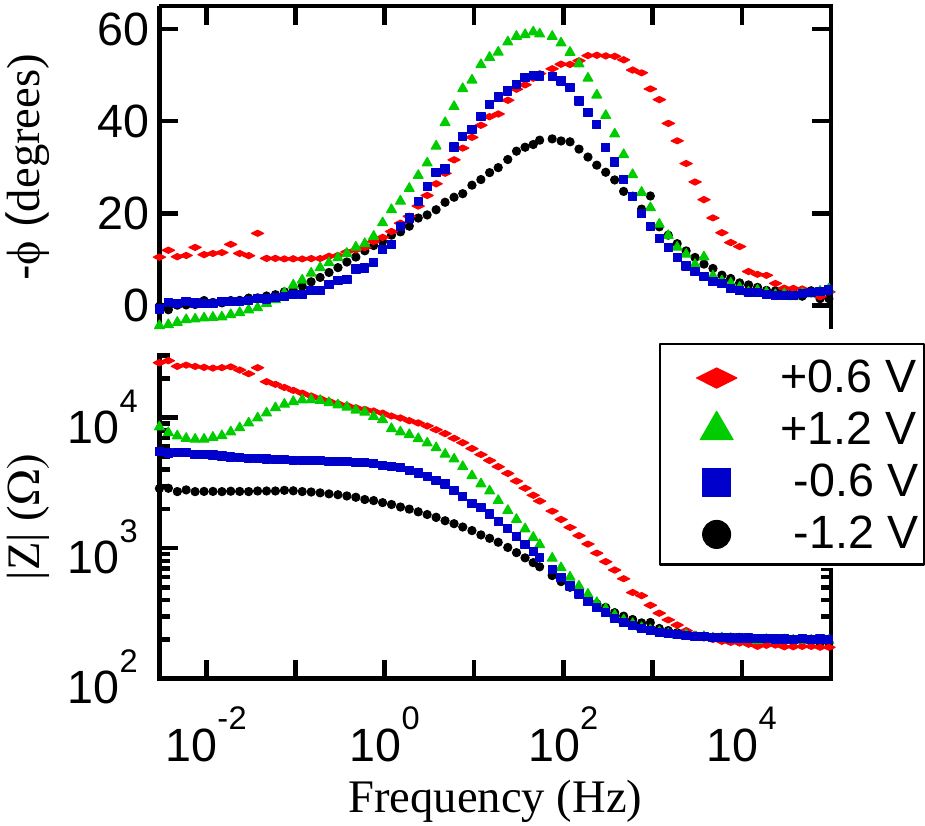}}
  \caption{Bode plot from electrochemical impedance spectroscopy (EIS) showing the phase angle and impedance magnitude vs. frequency. Data obtained concurrent to the APXPS measurements of overpotentials.}
  \label{fig_eis}
\end{figure}

Table~\ref{table_values} contains all the measured overpotentials and other relevant quantities obtained from the potentiostat and from the XPS data. The sum of all three total overpotentials is $\sim{50}$~mV lower than $V_{cell}$, likely due to the resistance of the cables and the contact resistance between the spring-loaded probes and the electrodes. The closeness of the overpotential sum to $V_{cell}$ shows that the measured overpotentials have internal consistency. An additional validity test comes from EIS measurements taken concurrently with the XPS data. Figure \ref{fig_eis}b shows a potentiostatic Bode plot. As the phase angle goes to zero at high-frequency, the impedance measures the global electrolyte resistance of the SOEC\cite{orazem_electrochemical_2008}, giving $R_{YSZ}=195\pm{20}~\Omega$. Dividing the APXPS-derived overpotential $\eta_{_{\rm{R}}}$ by $I_{cell}$ is an alternative way of calculating the electrolyte resistance. As shown in Table~\ref{table_values} the agreement between the two methods is excellent. In addition, this comparison shows that the current in the SOEC is spatially uniform because the value probed by XPS in a localized region agrees with the average full-cell (global) value. These two independent tests validate the methodology of using APXPS as a non-contact method of measuring overpotentials \textit{in~operando}. 

Figure \ref{fig_potentials}b plots the measured overpotentials for several  $V_{cell}$ values. The marker size is proportional to $I_{cell}$ (see Table~\ref{table_values}). We next discuss what information can be learned from how the individual overpotentials change with $V_{cell}$. At a given $V_{cell}$ value, both electrodes experience the same current ($I_{cell}$), have the same geometry and, thus, the same TPB length. $I_{cell}$ is therefore proportional to both current densities, $i_{\rm{_{Ni-YSZ}}}$ and $i_{\rm{_{YSZ-Pt}}}$. In addition, the electrodes experience the same temperature and the same gas composition and pressure. However, the electrodes have different steady-state overpotentials when current flows, as can be seen in Fig.~\ref{fig_potentials}b. The Butler-Volmer relationship (Eq.~\ref{eq:butler_volmer}) tells us that electrodes experiencing the same current densities ($i_{\rm{_{Ni-YSZ}}}=i_{\rm{_{YSZ-Pt}}}$) with different overpotentials must have different exchange-current densities ($i_{\rm{0,_{Ni-YSZ}}}{\neq}i_{\rm{0,_{YSZ-Pt}}}$). That is, the electrodes differ in their efficiency in performing the corresponding charge-transfer reaction. We next explore how the exchange-current densities of Pt and Ni compare for the two different cell reactions, H$_2$O splitting and H$_2$ oxidation (Eq.~\ref{eq:overall_rxn}).

The Ni overpotentials are always larger than the Pt overpotentials, whether Ni is splitting H$_2$O or oxidizing H$_2$, i.e., for $V_{cell} {<}$ 0 and $V_{cell} {>}$0, respectively (Fig.~\ref{fig_potentials}b). With the same steady-state current flowing through both electrodes, Eq.~\ref{eq:butler_volmer} then informs us that the larger overpotential of Ni occurs because its exchange-current density is smaller than that of Pt. This must be the case for both electrochemical reactions: 
\begin{equation}
{i}_{\rm{0,_{Ni-YSZ}}}^{WSR}{<}{i}_{\rm{0,_{YSZ-Pt}}}^{HOR} \;\;\;\; \text{and} \;\;\;\;
{i}_{\rm{0,_{Ni-YSZ}}}^{HOR}{<}{i}_{\rm{0,_{YSZ-Pt}}}^{WSR}
\label{eq:exchange}
\end{equation}
where the superscripts $WSR$ and $HOR$ refer to H$_2$O splitting and H$_2$ oxidation reactions, respectively. In other words, Ni needs a bigger fraction of $V_{cell}$ than Pt to compensate for its poor charge-transfer reaction rate. The higher exchange-current density of Pt is indicative of its superior electro-catalytic activity, i.e., it is more efficient for catalytic promotion of the charge-transfer reaction. Other contributions might include differences in the micro-structural quality of the electrodes and the accumulation of impurities at the TPB. While the performance difference between Ni and Pt is not at all surprising\cite{rossmeisl_trends_2008,li_electrocatalytic_2008}, we emphasize the directness of our approach to make such determinations.

Comparing the velocities of the H$_2$O splitting and H$_2$ oxidation reactions on the same electrode but at different values of $V_{cell}$ requires accounting for the cell current. We do this by considering the ratio between the overpotential and the current, the faradaic resistance $R_{F}$=$\eta$/$I_{cell}$, which represents the total reaction resistance. Lower faradaic resistance indicates a faster reaction velocity. When $V_{cell}$=${-0.6}$~V and $V_{cell}$= 0.6~V, Ni is performing $WSR$ and $HOR$, respectively, and $R_{F}(Ni,WSR,{-0.6}~V)=3.7{\times}10^{-3}$~$\Omega$ and $R_{F}(Ni,HOR,0.6~V) = 6.7{\times}10^{-3}$~$\Omega$. This comparison shows that H$_2$O splitting on Ni is faster than H$_2$ oxidation. At the same cell voltages, Pt is performing $HOR$ and $WSR$, respectively, and $R_{F}(Pt, HOR,{-0.6}~V)=2.3{\times}10^{-3}$~$\Omega$ and $R_{F}(Pt,WSR,0.6~V)= 3.5{\times}10^{-3}$~$\Omega$. Therefore, H$_2$ oxidation proceeds more rapidly than H$_2$O splitting on Pt.

The cell behaviour at $V_{cell}$=${-0.6}$ and {0.6}~V directly illustrates the importance of the electro-catalyst in the overall SOEC performance -- the Ni and Pt electrodes have similar overpotentials but the current is two times greater when hydrogen oxidation occurs on Pt ($V_{cell}$ = ${-0.6}$~V).

At $V_{cell}$=1.2~V, $\eta_{_{\rm{Ni-YSZ}}}$=1.02~V and $I_{cell}$=32~$\mu$A. At the same time, $\eta_{_{\rm{YSZ-Pt}}}$ is smaller than at $V_{cell}$=0.6~V. These observations are explained by the known\cite{chigane_xrd_1998,garcia-miquel_nickel_2003} interfacial oxidation of Ni, which occurs when $\eta_{_{\rm{Ni-YSZ}}}>{0.4}~V$. The oxide layer at the interface and the TPB greatly decreases the electro-catalytic activity of the Ni electrode, reducing ${i}_{\rm{0,_{Ni-YSZ}}}$ and consequently increasing $\eta_{\rm{bias,_{Ni-YSZ}}}$ and the total Ni overpotential, $\eta_{\rm{_{Ni-YSZ}}}$. Commercial fuel-cells avoid the oxidation of Ni by using porous Ni-YSZ cermets that increase the extent of the Ni electrode TPB and reduce the local current densities, also reducing $\eta_{_{\rm{Ni-YSZ}}}$. Because the current is smaller at {1.2}~V than at {0.6}~V, $\eta_{_{\rm{YSZ-Pt}}}$ is also smaller. The overpotential at the Ni/YSZ interface at $V_{cell}$=0.6 ($\eta_{_{\rm{Ni-YSZ}}}$=0.37~V) indicates that the Ni interface was not oxidized during these conditions.

\section{Conclusions}
\label{sec:conclusions}
We have demonstrated that photo-electron spectroscopy performed \textit{in~operando} at near ambient pressure can measure directly the individual overpotentials in solid-state electrochemical devices. In our simple electrically driven SOEC, the Ni and Pt electrodes are at the same temperature, see the same gas, and have the same symmetric geometry on the YSZ electrolyte. We can then interpret the origins of the changes in the individual overpotentials ($\eta_{_{\rm{Ni-YSZ}}}$, $\eta_{_{\rm{YSZ-Pt}}}$ and $\eta_{_{\rm{R}}}$) with applied bias in terms of the different electro-catalytic activities of Ni and Pt for the H$_2$O splitting and H$_2$ oxidation reactions. We find without ambiguity that H$_2$O splitting is faster than H$_2$ oxidation on Ni, while on Pt $H_2$ oxidation reaction proceeds more rapidly than $H_2$O splitting. In addition, by measuring the portion of the applied bias $V_{cell}$ that is consumed at the Ni/YSZ interface, $\eta_{_{\rm{Ni-YSZ}}}$, we can determine if the Ni at that interface is oxidized under different conditions.

Using photo-electrons as non-contact potential probes \textit{in~operando} allows direct access to all the individual overpotentials of an electrochemical cell. Knowing the ``potential landscape,'' along with the ``composition/chemical state landscape,'' should provide a deeper understanding of the basic mechanisms of fuel-cell and electrolyzer operation and enable improved performance.

\section{Acknowledgments}
\label{ack}
We thank C. Zhang, S. C. DeCaluwe, and B. W. Eichhorn for stimulating discussions, and G. S. Jackson (U. of Maryland) for comments on this manuscript. This research was supported by the U. S. Department of Energy through the Sandia Laboratory Directed Research and Development program under Contract DE-AC04-94AL85000. The Advanced Light Source is supported by the Director, Office of Science, Office of Basic Energy Sciences, of the U.S. Department of Energy under Contract No. DE-AC02-05CH11231.

\section*{Appendix}
\appendix

\section{SOEC potentials probed by photo-electrons} 
\label{potentials_XPS}
In this appendix, we provide a formal description of how XPS can directly measure the difference between biased and non-biased (equilibrium) states in a material's inner potential, ${}^{\rm{bias}}\Delta^{\rm{eq}}\phi$. The fundamental XPS process is the emission of a photo-electron from an atomic energy level (core or valence band) excited by monochromatic x-rays\cite{hufner_photoelectron_2003,siegbahn_x-ray_2008}. The emitted photo-electron will leave the surface if the kinetic energy it acquired from the incident x-ray photon, $E_{\rm{K.E.}}^{\rm{init}}$, can overcome all the potentials that bind it to the solid.

The electron kinetic energy measured at the spectrometer can be considered as the sum of discrete contributions. First, the electron has to overcome the atomic binding energy (referenced to the material's Fermi level), E$_{\rm{B.E.}}^f$, to escape from the atom. E$_{\rm{B.E.}}^f$ is defined as the energy difference between the initial, unperturbed N electron atom and the final-state {N-1} electron atom. After escaping the atom, the electron has to travel through the solid toward the surface. Since we are only interested in the elastic electrons present in XPS peaks and not the broad inelastic backgrounds, we have to consider the electron mean free path in the solid. For the kinetic energies used in this work, the mean free path is $\sim$1~nm, i.e., electrons created within $\sim$1~nm of the surface will escape elastically out of the solid. 

The second term to consider is the chemical potential of the electron, $\mu_{\rm{e}}$, which is the negative of the energy cost of removing one electron at the Fermi energy from the solid. ($\mu_{\rm{e}}$ is defined as introducing one electron into the material.) This energy is characteristic of a material. The third contribution arises if the sample and the spectrometer are not Fermi-edge coupled\cite{ebel_absolute_1976,lewis_binding-energy_1980,barr_modern_1994,siegbahn_x-ray_2008}. Then the difference between their Fermi levels will contribute to the measured kinetic energy. We will call this term the Fermi-level difference, $\delta$F$_{\rm{E}}$. The fourth contribution is the energy cost of crossing the  surface potential $\chi$, which results in an energy of $\rm{e}\chi$, where $\rm{e}$ is the charge of the electron. The combination of the electron chemical potential with negative sign and the energy change in crossing the surface potential dipole $\chi$ is the electron work function\cite{van_rysselberghe_notework_1953,holzl_work_1979,trasatti_interphases_1986}: $\Phi_{\rm{sample}}=-\mu_{\rm{e}}+\rm{e}\chi$. The fifth (last) contribution comes from the spectrometer ---~the electron has to overcome the spectrometer work function ($\Phi_{\rm{sp}}$) to enter and be counted.

When a driven cell is measured under bias, double-layers will form and overpotentials will develop at both electrode/electrolyte interfaces (see Section~\ref{potentials}). Since all the other contributions to the photo-electron kinetic energy just described (E$_{\rm{B.E.}}^f$, $\mu_{\rm{e}}$, $\rm{e}\chi$\cite{foot2} and $\Phi_{\rm{sp}}$) are independent of the bias, the difference in the kinetic energies of photo-electrons from the same region with applied bias $E_{\rm{K.E.}}^{\rm{bias}}$ and with no external bias $E_{\rm{K.E.}}^{\rm{eq}}$ measures the change in the Fermi level difference\cite{edgell_biased_1986,ahn_abnormal_2006}:
\begin{equation}
E_{\rm{K.E.}}^{\rm{bias}} - E_{\rm{K.E.}}^{\rm{eq}} = {}^{\rm{bias}}\Delta^{\rm{eq}}\delta\rm{F}_{\rm{E}}.
\end{equation}
Only the Ni electrode is Fermi-edge coupled to the XPS spectrometer, so $\delta\rm{F}_{\rm{E}}^{Ni}=0$ for all bias conditions. When bias is applied between Ni and Pt electrodes, the Fermi level of the YSZ and Pt will change with respect to the XPS spectrometer ($\delta\rm{F}_{\rm{E}} {\neq} 0$). To relate the Fermi energy and the inner potential, we recall that the Fermi level of a particular phase, $\rm{F}_{\rm{E}}$, corresponds\cite{bockris_fermi_1983,bard_electrochemical_2001} to the electrochemical potential of its electrons, $\overline{\mu}_{\rm{e}}$:
\begin{equation}
\rm{F}_{\rm{E}} = \overline{\mu}_{\rm{e}} = \mu_{\rm{e}} - \rm{e}\phi.
\end{equation}
Since $\mu_{\rm{e}}$ does not depend on bias, the bias-induced change in $\delta\rm{F}_{\rm{E}}$ corresponds to:
\begin{equation}
E_{\rm{K.E.}}^{\rm{bias}} - E_{\rm{K.E.}}^{\rm{eq}} = -{\rm{e}}^{\rm{bias}}\Delta^{\rm{eq}}{\phi}.
\label{eq:delta_inner_apend_eV}
\end{equation}
If the kinetic energy is measured in eV, $\rm{e}=1$~{eV/V} and:
\begin{equation}
E_{\rm{K.E.}}^{\rm{bias}} - E_{\rm{K.E.}}^{\rm{eq}} = -{}^{\rm{bias}}\Delta^{\rm{eq}}\phi   \;\;\;\;\;\;\text{(in eV)}.
\label{eq:delta_inner_apend_V}
\end{equation}
That is, by measuring how the kinetic energies of the XPS peaks shift rigidly with bias, the change in inner potential can be measured\cite{ladas_origin_1993}. In Section~\ref{results} we show how this relationship can be combined with the equations in Section~\ref{potentials} to directly measure the total overpotential of a double-layer. We emphasize the key point that the photo-electrons have finite escape depth, $\sim$1~nm in our experiment. Thus, we are measuring the inner potential changes in the material's near-surface region. In contrast, the surface potential arises from the atomically-sharp discontinuity in the material at its surface.



\renewcommand\refname{Notes and references}

\footnotesize{
\bibliography{ElGabaly_potentials} 

\providecommand*{\mcitethebibliography}{\thebibliography}
\csname @ifundefined\endcsname{endmcitethebibliography}
{\let\endmcitethebibliography\endthebibliography}{}
\begin{mcitethebibliography}{46}
\providecommand*{\natexlab}[1]{#1}
\providecommand*{\mciteSetBstSublistMode}[1]{}
\providecommand*{\mciteSetBstMaxWidthForm}[2]{}
\providecommand*{\mciteBstWouldAddEndPuncttrue}
  {\def\EndOfBibitem{\unskip.}}
\providecommand*{\mciteBstWouldAddEndPunctfalse}
  {\let\EndOfBibitem\relax}
\providecommand*{\mciteSetBstMidEndSepPunct}[3]{}
\providecommand*{\mciteSetBstSublistLabelBeginEnd}[3]{}
\providecommand*{\EndOfBibitem}{}
\mciteSetBstSublistMode{f}
\mciteSetBstMaxWidthForm{subitem}
{(\emph{\alph{mcitesubitemcount}})}
\mciteSetBstSublistLabelBeginEnd{\mcitemaxwidthsubitemform\space}
{\relax}{\relax}

\bibitem[Winter and Brodd(2004)]{winter_what_2004}
M.~Winter and R.~J. Brodd, \emph{Chemical Reviews}, 2004, \textbf{104},
  4245--4270\relax
\mciteBstWouldAddEndPuncttrue
\mciteSetBstMidEndSepPunct{\mcitedefaultmidpunct}
{\mcitedefaultendpunct}{\mcitedefaultseppunct}\relax
\EndOfBibitem
\bibitem[Haile(2003)]{haile_fuel_2003}
S.~M. Haile, \emph{Acta Materialia}, 2003, \textbf{51}, 5981--6000\relax
\mciteBstWouldAddEndPuncttrue
\mciteSetBstMidEndSepPunct{\mcitedefaultmidpunct}
{\mcitedefaultendpunct}{\mcitedefaultseppunct}\relax
\EndOfBibitem
\bibitem[Kirubakaran \emph{et~al.}(2009)Kirubakaran, Jain, and
  Nema]{kirubakaran_reviewfuel_2009}
A.~Kirubakaran, S.~Jain and R.~Nema, \emph{Renewable and Sustainable Energy
  Reviews}, 2009, \textbf{13}, 2430--2440\relax
\mciteBstWouldAddEndPuncttrue
\mciteSetBstMidEndSepPunct{\mcitedefaultmidpunct}
{\mcitedefaultendpunct}{\mcitedefaultseppunct}\relax
\EndOfBibitem
\bibitem[Ni \emph{et~al.}(2008)Ni, Leung, and Leung]{ni_technological_2008}
M.~Ni, M.~K. Leung and D.~Y. Leung, \emph{International Journal of Hydrogen
  Energy}, 2008, \textbf{33}, 2337--2354\relax
\mciteBstWouldAddEndPuncttrue
\mciteSetBstMidEndSepPunct{\mcitedefaultmidpunct}
{\mcitedefaultendpunct}{\mcitedefaultseppunct}\relax
\EndOfBibitem
\bibitem[Grove(1839)]{grove_xxiv.voltaic_1839}
W.~R. Grove, \emph{Philosophical Magazine Series 3}, 1839, \textbf{14},
  127\relax
\mciteBstWouldAddEndPuncttrue
\mciteSetBstMidEndSepPunct{\mcitedefaultmidpunct}
{\mcitedefaultendpunct}{\mcitedefaultseppunct}\relax
\EndOfBibitem
\bibitem[Appleby(1990)]{appleby_sir_1990}
A.~J. Appleby, \emph{Journal of Power Sources}, 1990, \textbf{29}, 3--11\relax
\mciteBstWouldAddEndPuncttrue
\mciteSetBstMidEndSepPunct{\mcitedefaultmidpunct}
{\mcitedefaultendpunct}{\mcitedefaultseppunct}\relax
\EndOfBibitem
\bibitem[And\'{u}jar and Segura(2009)]{andjar_fuel_2009}
J.~And\'{u}jar and F.~Segura, \emph{Renewable and Sustainable Energy Reviews},
  2009, \textbf{13}, 2309--2322\relax
\mciteBstWouldAddEndPuncttrue
\mciteSetBstMidEndSepPunct{\mcitedefaultmidpunct}
{\mcitedefaultendpunct}{\mcitedefaultseppunct}\relax
\EndOfBibitem
\bibitem[Vogler \emph{et~al.}(2009)Vogler, {Bieberle-Hutter}, Gauckler,
  Warnatz, and Bessler]{vogler_modelling_2009}
M.~Vogler, A.~{Bieberle-Hutter}, L.~Gauckler, J.~Warnatz and W.~G. Bessler,
  \emph{Journal of The Electrochemical Society}, 2009, \textbf{156},
  B663--B672\relax
\mciteBstWouldAddEndPuncttrue
\mciteSetBstMidEndSepPunct{\mcitedefaultmidpunct}
{\mcitedefaultendpunct}{\mcitedefaultseppunct}\relax
\EndOfBibitem
\bibitem[Goodwin \emph{et~al.}(2009)Goodwin, Zhu, Colclasure, and
  Kee]{goodwin_modeling_2009}
D.~G. Goodwin, H.~Zhu, A.~M. Colclasure and R.~J. Kee, \emph{Journal of The
  Electrochemical Society}, 2009, \textbf{156}, B1004--B1021\relax
\mciteBstWouldAddEndPuncttrue
\mciteSetBstMidEndSepPunct{\mcitedefaultmidpunct}
{\mcitedefaultendpunct}{\mcitedefaultseppunct}\relax
\EndOfBibitem
\bibitem[Bockris \emph{et~al.}(2000)Bockris, Reddy, and
  {Gamboa-Aldeco}]{bockris_modern_2000}
J.~O. Bockris, A.~K.~N. Reddy and M.~{Gamboa-Aldeco}, \emph{Modern
  Electrochemistry {2A}}, Springer, Berlin, 2000\relax
\mciteBstWouldAddEndPuncttrue
\mciteSetBstMidEndSepPunct{\mcitedefaultmidpunct}
{\mcitedefaultendpunct}{\mcitedefaultseppunct}\relax
\EndOfBibitem
\bibitem[foo()]{foot3}
For example, the charge time of a Li-ion battery is much slower than the
  discharge because of large overpotentials when the current is flowing in the
  charging direction.\relax
\mciteBstWouldAddEndPunctfalse
\mciteSetBstMidEndSepPunct{\mcitedefaultmidpunct}
{}{\mcitedefaultseppunct}\relax
\EndOfBibitem
\bibitem[N\"{o}rskov \emph{et~al.}(2004)N\"{o}rskov, Rossmeisl, Logadottir,
  Lindqvist, Kitchin, Bligaard, and J\'{o}nsson]{nrskov_origin_2004}
J.~K. N\"{o}rskov, J.~Rossmeisl, A.~Logadottir, L.~Lindqvist, J.~R. Kitchin,
  T.~Bligaard and H.~J\'{o}nsson, \emph{The Journal of Physical Chemistry B},
  2004, \textbf{108}, 17886--17892\relax
\mciteBstWouldAddEndPuncttrue
\mciteSetBstMidEndSepPunct{\mcitedefaultmidpunct}
{\mcitedefaultendpunct}{\mcitedefaultseppunct}\relax
\EndOfBibitem
\bibitem[Mann \emph{et~al.}(2006)Mann, Amphlett, Peppley, and
  Thurgood]{mann_application_2006}
R.~Mann, J.~Amphlett, B.~Peppley and C.~Thurgood, \emph{Journal of Power
  Sources}, 2006, \textbf{161}, 775--781\relax
\mciteBstWouldAddEndPuncttrue
\mciteSetBstMidEndSepPunct{\mcitedefaultmidpunct}
{\mcitedefaultendpunct}{\mcitedefaultseppunct}\relax
\EndOfBibitem
\bibitem[Bessler \emph{et~al.}(2007)Bessler, Warnatz, and
  Goodwin]{bessler_influence_2007}
W.~G. Bessler, J.~Warnatz and D.~G. Goodwin, \emph{Solid State Ionics}, 2007,
  \textbf{177}, 3371--3383\relax
\mciteBstWouldAddEndPuncttrue
\mciteSetBstMidEndSepPunct{\mcitedefaultmidpunct}
{\mcitedefaultendpunct}{\mcitedefaultseppunct}\relax
\EndOfBibitem
\bibitem[Gileadi(1993)]{gileadi_electrode_1993}
E.~Gileadi, \emph{Electrode Kinetics for Chemists, Chemical Engineers and
  Materials Scientists}, John Wiley and Sons, New York, 1993\relax
\mciteBstWouldAddEndPuncttrue
\mciteSetBstMidEndSepPunct{\mcitedefaultmidpunct}
{\mcitedefaultendpunct}{\mcitedefaultseppunct}\relax
\EndOfBibitem
\bibitem[Ogletree \emph{et~al.}(2002)Ogletree, Bluhm, Lebedev, Fadley, Hussain,
  and Salmeron]{ogletree_differentially_2002}
D.~F. Ogletree, H.~Bluhm, G.~Lebedev, C.~S. Fadley, Z.~Hussain and M.~Salmeron,
  \emph{Review of Scientific Instruments}, 2002, \textbf{73}, 3872--3877\relax
\mciteBstWouldAddEndPuncttrue
\mciteSetBstMidEndSepPunct{\mcitedefaultmidpunct}
{\mcitedefaultendpunct}{\mcitedefaultseppunct}\relax
\EndOfBibitem
\bibitem[Salmeron and Schl\"{o}gl(2008)]{salmeron_ambient_2008}
M.~Salmeron and R.~Schl\"{o}gl, \emph{Surface Science Reports}, 2008,
  \textbf{63}, 169--199\relax
\mciteBstWouldAddEndPuncttrue
\mciteSetBstMidEndSepPunct{\mcitedefaultmidpunct}
{\mcitedefaultendpunct}{\mcitedefaultseppunct}\relax
\EndOfBibitem
\bibitem[Grass \emph{et~al.}(2010)Grass, Karlsson, Aksoy, M.~Lundqvist, Mun,
  Hussain, and Liu.]{Grass_2010}
M.~E. Grass, P.~G. Karlsson, F.~Aksoy, B.~W. M.~Lundqvist, B.~S. Mun,
  Z.~Hussain and Z.~Liu., \textit{Review of Scientific Instruments},
  submitted\relax
\mciteBstWouldAddEndPuncttrue
\mciteSetBstMidEndSepPunct{\mcitedefaultmidpunct}
{\mcitedefaultendpunct}{\mcitedefaultseppunct}\relax
\EndOfBibitem
\bibitem[Fahlman \emph{et~al.}(1966)Fahlman, Hamrin, Hedman, Nordberg,
  Nordling, and Siegbahn]{fahlman_electron_1966}
A.~Fahlman, K.~Hamrin, J.~Hedman, R.~Nordberg, C.~Nordling and K.~Siegbahn,
  \emph{Nature}, 1966, \textbf{210}, 4--8\relax
\mciteBstWouldAddEndPuncttrue
\mciteSetBstMidEndSepPunct{\mcitedefaultmidpunct}
{\mcitedefaultendpunct}{\mcitedefaultseppunct}\relax
\EndOfBibitem
\bibitem[Siegbahn and Lundholm(1982)]{siegbahn_method_1982}
H.~Siegbahn and M.~Lundholm, \emph{Journal of Electron Spectroscopy and Related
  Phenomena}, 1982, \textbf{28}, 135--138\relax
\mciteBstWouldAddEndPuncttrue
\mciteSetBstMidEndSepPunct{\mcitedefaultmidpunct}
{\mcitedefaultendpunct}{\mcitedefaultseppunct}\relax
\EndOfBibitem
\bibitem[Whaley(2010)]{Whaley_2010}
J.~Whaley, \textit{Manuscript in preparation}\relax
\mciteBstWouldAddEndPuncttrue
\mciteSetBstMidEndSepPunct{\mcitedefaultmidpunct}
{\mcitedefaultendpunct}{\mcitedefaultseppunct}\relax
\EndOfBibitem
\bibitem[foo()]{foot5}
The total pressure increases from 300~mTorr at room temperature to 380~mTorr at
  700~C.\relax
\mciteBstWouldAddEndPunctfalse
\mciteSetBstMidEndSepPunct{\mcitedefaultmidpunct}
{}{\mcitedefaultseppunct}\relax
\EndOfBibitem
\bibitem[Chiang(1997)]{chiang_physical_1997}
Y.~Chiang, \emph{Physical ceramics}, J. Wiley, New York, 1997\relax
\mciteBstWouldAddEndPuncttrue
\mciteSetBstMidEndSepPunct{\mcitedefaultmidpunct}
{\mcitedefaultendpunct}{\mcitedefaultseppunct}\relax
\EndOfBibitem
\bibitem[Bard(2001)]{bard_electrochemical_2001}
A.~Bard, \emph{Electrochemical methods : fundamentals and applications}, Wiley,
  New York, 2nd edn, 2001\relax
\mciteBstWouldAddEndPuncttrue
\mciteSetBstMidEndSepPunct{\mcitedefaultmidpunct}
{\mcitedefaultendpunct}{\mcitedefaultseppunct}\relax
\EndOfBibitem
\bibitem[Trasatti and Parsons(1986)]{trasatti_interphases_1986}
S.~Trasatti and R.~Parsons, \emph{Pure and Applied Chemistry}, 1986,
  \textbf{58}, 437--454\relax
\mciteBstWouldAddEndPuncttrue
\mciteSetBstMidEndSepPunct{\mcitedefaultmidpunct}
{\mcitedefaultendpunct}{\mcitedefaultseppunct}\relax
\EndOfBibitem
\bibitem[Trasatti(1986)]{trasatti_absolute_1986}
S.~Trasatti, \emph{Pure and Applied Chemistry}, 1986, \textbf{58},
  955--966\relax
\mciteBstWouldAddEndPuncttrue
\mciteSetBstMidEndSepPunct{\mcitedefaultmidpunct}
{\mcitedefaultendpunct}{\mcitedefaultseppunct}\relax
\EndOfBibitem
\bibitem[Adler(2004)]{adler_factors_2004}
S.~B. Adler, \emph{Chemical Reviews}, 2004, \textbf{104}, 4791--4844\relax
\mciteBstWouldAddEndPuncttrue
\mciteSetBstMidEndSepPunct{\mcitedefaultmidpunct}
{\mcitedefaultendpunct}{\mcitedefaultseppunct}\relax
\EndOfBibitem
\bibitem[Noren and Hoffman(2005)]{noren_clarifyingbutler-volmer_2005}
D.~Noren and M.~Hoffman, \emph{Journal of Power Sources}, 2005, \textbf{152},
  175--181\relax
\mciteBstWouldAddEndPuncttrue
\mciteSetBstMidEndSepPunct{\mcitedefaultmidpunct}
{\mcitedefaultendpunct}{\mcitedefaultseppunct}\relax
\EndOfBibitem
\bibitem[foo()]{foot4}
A critical advantage of our spectral imaging approach is that we can
  distinguish electrode material that becomes electrically disconnected from
  the electrode. Such material can occur at the metal/electrolyte boundary
  where the metal de-wets the substrate, forming isolated metal islands. Using
  the element-specific and spatially resolved PES peaks, we can directly
  distinguish metal islands that are not at the electrode potential but at the
  same potential as the surrounding YSZ.\relax
\mciteBstWouldAddEndPunctfalse
\mciteSetBstMidEndSepPunct{\mcitedefaultmidpunct}
{}{\mcitedefaultseppunct}\relax
\EndOfBibitem
\bibitem[Orazem(2008)]{orazem_electrochemical_2008}
M.~Orazem, \emph{Electrochemical impedance spectroscopy}, Wiley, Hoboken
  {N.J.}, 2008\relax
\mciteBstWouldAddEndPuncttrue
\mciteSetBstMidEndSepPunct{\mcitedefaultmidpunct}
{\mcitedefaultendpunct}{\mcitedefaultseppunct}\relax
\EndOfBibitem
\bibitem[Rossmeisl and Bessler(2008)]{rossmeisl_trends_2008}
J.~Rossmeisl and W.~G. Bessler, \emph{Solid State Ionics}, 2008, \textbf{178},
  1694--1700\relax
\mciteBstWouldAddEndPuncttrue
\mciteSetBstMidEndSepPunct{\mcitedefaultmidpunct}
{\mcitedefaultendpunct}{\mcitedefaultseppunct}\relax
\EndOfBibitem
\bibitem[Li \emph{et~al.}(2008)Li, Lee, and Zhang]{li_electrocatalytic_2008}
H.~Li, K.~Lee and J.~Zhang, \emph{{PEM} Fuel Cell Electrocatalysts and Catalyst
  Layers}, 2008, pp. 135--164\relax
\mciteBstWouldAddEndPuncttrue
\mciteSetBstMidEndSepPunct{\mcitedefaultmidpunct}
{\mcitedefaultendpunct}{\mcitedefaultseppunct}\relax
\EndOfBibitem
\bibitem[Chigane and Ishikawa(1998)]{chigane_xrd_1998}
M.~Chigane and M.~Ishikawa, \emph{Journal of the Chemical Society, Faraday
  Transactions}, 1998, \textbf{94}, 3665--3670\relax
\mciteBstWouldAddEndPuncttrue
\mciteSetBstMidEndSepPunct{\mcitedefaultmidpunct}
{\mcitedefaultendpunct}{\mcitedefaultseppunct}\relax
\EndOfBibitem
\bibitem[{Garcia-Miquel} \emph{et~al.}(2003){Garcia-Miquel}, Zhang, Allen,
  Rougier, Blyr, Davies, Jones, Leedham, Williams, and
  Impey]{garcia-miquel_nickel_2003}
J.~L. {Garcia-Miquel}, Q.~Zhang, S.~J. Allen, A.~Rougier, A.~Blyr, H.~O.
  Davies, A.~C. Jones, T.~J. Leedham, P.~A. Williams and S.~A. Impey,
  \emph{Thin Solid Films}, 2003, \textbf{424}, 165--170\relax
\mciteBstWouldAddEndPuncttrue
\mciteSetBstMidEndSepPunct{\mcitedefaultmidpunct}
{\mcitedefaultendpunct}{\mcitedefaultseppunct}\relax
\EndOfBibitem
\bibitem[H\"{u}fner(2003)]{hufner_photoelectron_2003}
S.~H\"{u}fner, \emph{Photoelectron spectroscopy: principles and applications},
  Springer, Berlin, 3rd edn, 2003\relax
\mciteBstWouldAddEndPuncttrue
\mciteSetBstMidEndSepPunct{\mcitedefaultmidpunct}
{\mcitedefaultendpunct}{\mcitedefaultseppunct}\relax
\EndOfBibitem
\bibitem[Siegbahn(2008)]{siegbahn_x-ray_2008}
K.~Siegbahn, \emph{Nishina Memorial Lectures}, Springer Japan, Tokyo, 2008,
  vol. 746, pp. 137--228\relax
\mciteBstWouldAddEndPuncttrue
\mciteSetBstMidEndSepPunct{\mcitedefaultmidpunct}
{\mcitedefaultendpunct}{\mcitedefaultseppunct}\relax
\EndOfBibitem
\bibitem[Ebel(1976)]{ebel_absolute_1976}
M.~F. Ebel, \emph{Journal of Electron Spectroscopy and Related Phenomena},
  1976, \textbf{8}, 213--224\relax
\mciteBstWouldAddEndPuncttrue
\mciteSetBstMidEndSepPunct{\mcitedefaultmidpunct}
{\mcitedefaultendpunct}{\mcitedefaultseppunct}\relax
\EndOfBibitem
\bibitem[Lewis and Kelly(1980)]{lewis_binding-energy_1980}
R.~T. Lewis and M.~A. Kelly, \emph{Journal of Electron Spectroscopy and Related
  Phenomena}, 1980, \textbf{20}, 105--115\relax
\mciteBstWouldAddEndPuncttrue
\mciteSetBstMidEndSepPunct{\mcitedefaultmidpunct}
{\mcitedefaultendpunct}{\mcitedefaultseppunct}\relax
\EndOfBibitem
\bibitem[Barr(1994)]{barr_modern_1994}
T.~L. Barr, \emph{Modern {ESCA}}, {CRC} Press, Boca Raton, {FL}, 1994\relax
\mciteBstWouldAddEndPuncttrue
\mciteSetBstMidEndSepPunct{\mcitedefaultmidpunct}
{\mcitedefaultendpunct}{\mcitedefaultseppunct}\relax
\EndOfBibitem
\bibitem[Rysselberghe(1953)]{van_rysselberghe_notework_1953}
P.~V. Rysselberghe, \emph{The Journal of Chemical Physics}, 1953, \textbf{21},
  1550--1551\relax
\mciteBstWouldAddEndPuncttrue
\mciteSetBstMidEndSepPunct{\mcitedefaultmidpunct}
{\mcitedefaultendpunct}{\mcitedefaultseppunct}\relax
\EndOfBibitem
\bibitem[H\"{o}lzl and Schulte(1979)]{holzl_work_1979}
J.~H\"{o}lzl and F.~K. Schulte, \emph{Solid Surface Physics},
  {Springer-Verlag}, Berlin, 1979, vol.~85, pp. 1--150\relax
\mciteBstWouldAddEndPuncttrue
\mciteSetBstMidEndSepPunct{\mcitedefaultmidpunct}
{\mcitedefaultendpunct}{\mcitedefaultseppunct}\relax
\EndOfBibitem
\bibitem[foo()]{foot2}
Although the surface potential, $\chi$, will change if the surface adsorbates
  change with external bias, we should note that the surfaces probed by PES in
  this experiment (electrode-gas and electrolyte-gas, A and B in
  Fig.~\ref{fig_double}, respectively) do not have double-layers and
  consequently, do not change their near-surface electric field. As a
  consequence we do not expect adsorbates and $\chi$ to change significantly
  with the bias.\relax
\mciteBstWouldAddEndPunctfalse
\mciteSetBstMidEndSepPunct{\mcitedefaultmidpunct}
{}{\mcitedefaultseppunct}\relax
\EndOfBibitem
\bibitem[Edgell \emph{et~al.}(1986)Edgell, Baer, and
  Castle]{edgell_biased_1986}
M.~Edgell, D.~Baer and J.~Castle, \emph{Applied Surface Science}, 1986,
  \textbf{26}, 129--149\relax
\mciteBstWouldAddEndPuncttrue
\mciteSetBstMidEndSepPunct{\mcitedefaultmidpunct}
{\mcitedefaultendpunct}{\mcitedefaultseppunct}\relax
\EndOfBibitem
\bibitem[Ahn \emph{et~al.}(2006)Ahn, Zharnikov, and Whitten]{ahn_abnormal_2006}
H.~Ahn, M.~Zharnikov and J.~E. Whitten, \emph{Chemical Physics Letters}, 2006,
  \textbf{428}, 283--287\relax
\mciteBstWouldAddEndPuncttrue
\mciteSetBstMidEndSepPunct{\mcitedefaultmidpunct}
{\mcitedefaultendpunct}{\mcitedefaultseppunct}\relax
\EndOfBibitem
\bibitem[Bockris and Khan(1983)]{bockris_fermi_1983}
J.~O. Bockris and S.~U.~M. Khan, \emph{Applied Physics Letters}, 1983,
  \textbf{42}, 124--125\relax
\mciteBstWouldAddEndPuncttrue
\mciteSetBstMidEndSepPunct{\mcitedefaultmidpunct}
{\mcitedefaultendpunct}{\mcitedefaultseppunct}\relax
\EndOfBibitem
\bibitem[Ladas \emph{et~al.}(1993)Ladas, Kennou, Bebelis, and
  Vayenas]{ladas_origin_1993}
S.~Ladas, S.~Kennou, S.~Bebelis and C.~G. Vayenas, \emph{The Journal of
  Physical Chemistry}, 1993, \textbf{97}, 8845--8848\relax
\mciteBstWouldAddEndPuncttrue
\mciteSetBstMidEndSepPunct{\mcitedefaultmidpunct}
{\mcitedefaultendpunct}{\mcitedefaultseppunct}\relax
\EndOfBibitem
\end{mcitethebibliography}
\bibliographystyle{rsc} 
}

\end{document}